\documentclass[aps,physrev,reprint,floatfix,superscriptaddress]{revtex4-2}

\usepackage[T1]{fontenc}
\usepackage[utf8]{inputenc}
\usepackage{verbatim}
\usepackage{amsmath, amssymb, amsfonts, bm, amsthm, braket}
\usepackage{graphicx}
\usepackage{color}
\usepackage{hyperref} % Must always be last

\newcommand{\code}{\texttt}
\newcommand{\norm}[1]{\left\lVert#1\right\rVert}

\newcommand{\Google}{\affiliation{%
Google Quantum AI, Mountain View, California 94043, United States}}

\newcommand{\QSimulate}{\affiliation{%
Quantum Simulation Technologies Inc., Boston, Massachusetts 02135, United States}}

\newcommand{\caltechchem}{\affiliation{%
Division of Chemistry and Chemical Engineering, California Institute of Technology, Pasadena, California 91125, United States}}

\date{November 2023}
\begin{document}
\title{Fast emulation of fermionic circuits with matrix product states}
\author{Justin Provazza}
\QSimulate

\author{Klaas Gunst}
\QSimulate

\author{Huanchen Zhai}
\caltechchem

\author{Garnet~K.-L.~Chan}
\caltechchem

\author{Toru Shiozaki}
\QSimulate

\author{Nicholas C.~Rubin}
\Google

\author{Alec F.~White}
\email{white@qsimulate.com}
\QSimulate
%\begin{document}
\begin{abstract}
We describe a matrix product state (MPS) extension for the Fermionic Quantum Emulator (FQE) software library. We discuss the theory behind symmetry adapted matrix product states for approximating many-body wavefunctions of spin-1/2 fermions, and we present an open-source, MPS-enabled implementation of the FQE interface (MPS-FQE). The software uses the open-source pyblock3 and block2 libraries for most elementary tensor operations, and it can largely be used as a drop-in replacement for FQE that allows for more efficient, but approximate, emulation of larger fermionic circuits. Finally, we show several applications relevant to both near-term and fault-tolerant quantum algorithms where approximate emulation of larger systems is expected to be useful: characterization of state preparation strategies for quantum phase estimation, the testing of different variational quantum eigensolver ans\"atze, the numerical evaluation of Trotter errors, and the simulation of general quantum dynamics problems. In all these examples, approximate emulation with MPS-FQE allows us to treat systems that are significantly larger than those accessible with a full statevector emulator.
\end{abstract}

\maketitle

\section{Introduction}
One of the most studied applications of quantum computers is fermionic simulation. In order to quantify the degree of computational advantage a quantum computation provides, the best classical strategies must be brought to bear on the equivalent problem. This has led to rapid development in classical strategies for approximating quantum circuits depending on the desired output from a circuit simulation--i.e. amplitudes, probabilities, or expectation values of operators. For example, approximate tensor network contraction~\cite{pan2020contracting, gray2021hyper, pan2022simulation, liu2021closing}, stabilizer simulation~\cite{bravyi2019simulation}, or Clifford perturbation theory~\cite{beguvsic2023simulating} have been used to classically emulate a quantum computational task at much lower complexity than 
exact state vector simulation 
%direct simulation 
of the %equivalent 
quantum circuit. Beyond defining the boundary of quantum advantage, efficient circuit simulation has become crucial for quantum algorithm design and analysis~\cite{stair2022qforte, nguyen2022tensor, strano2023exact, bayraktar2023cuquantum, zhang2023tensorcircuit}.  
In fact, the requirement for computational utility does not demand that classical emulators consider precisely the same circuit as the quantum implementation, so long as the same result is obtained. In fermionic settings, this has led to the development of fermion-specific simulators and frameworks that take advantage of additional structure in the problems of interest~\cite{rubin2021fermionic, stair2022qforte} resulting in substantially lower emulation costs.

For fermionic problems in particular, the Fermionic Quantum Emulator (FQE) is a software package that aims for maximum efficiency in full statevector emulation~\cite{rubin2021fermionic}. The FQE implements algorithms designed specifically to take full advantage of symmetries that are common to fermionic systems, in particular the particle number and projected spin symmetries. This source of efficiency enables emulation of quantum circuits that target fermionic simulation at larger scales than are achievable with naive statevector emulation. FQE provides near-optimal performance for most emulation tasks on single-node, CPU platforms. In a recent work~\cite{rubin2021fermionic}, it was shown that exploiting just number and projected spin-symmetries can lead to substantial savings in total run-time and memory requirements over qubit based simulations that do not use these symmetries.

Approximate approaches, like those based upon tensor networks (including matrix product states (MPS)) \cite{markov2008simulating, pan2020contracting, gray2021hyper, pan2022simulation, haghshenas2022variational, wahl2023simulating}, are instrumental for pushing the boundaries of classical emulation and have helped inform our understanding of what constitutes a classically intractable problem \cite{huang2020classical, ayral2207density, zhou2020limits, stoudenmire2023grover, chen2023quantum, beguvsic2023simulating, tindall2023efficient}. However, many implementations of these approximate approaches in the quantum emulation setting do not take advantage of problem symmetries \cite{cirq_developers_2021_4586899, Qiskit}.
In this work, we present a matrix product state backend for FQE. This software package implements the FQE interface, but uses an approximate MPS representation of the fermionic wavefunction instead of the exact statevector. The MPS implementation takes full advantage of particle number and projected spin symmetry, and the approximation is systematically improvable. The theory and applications presented here are specific to spin-1/2 fermions, but an extension to fermions with higher spin is straightforward.

The FQE simulator interfaces with the fermionic quantum simulation library OpenFermion \cite{mcclean2020openfermion}, as well as the more general quantum circuit emulator Cirq \cite{cirq_developers_2021_4586899}. The result is a powerful and convenient software ecosystem that enables highly efficient quantum emulation for a broad range of applications, and the MPS backend described here \cite{mps_fqe_2023} further expands the scope of treatable systems. In Section~\ref{sec:Theory} we briefly review the theory behind operations on MPS wavefunctions, in Section~\ref{sec:Implementation} we describe the software implementation, and in Section~\ref{sec:Applications} we show some example applications relevant to noisy intermediate-scale quantum (NISQ) and fault-tolerant quantum computing (FTQC) algorithms for quantum simulation.

\section{Theoretical Background}\label{sec:Theory}
The Hilbert space of many-particle systems grows exponentially with respect to their size. For spin-1/2 fermions, each orbital can be unoccupied, occupied by a single particle of up or down spin, or doubly occupied by particles of both spin: $\ket{\phi_i} \in [\ket{-}, \ket{\uparrow}, \ket{\downarrow}, \ket{\uparrow \downarrow}]$. This leads to 4 possible occupations for each orbital and $4^N$ many-particle basis states for $N$ spatial orbitals, and we refer to this as the standard fermionic representation. If one maps the problem to qubits for the purpose of simulation on a quantum computer, then the Jordan--Wigner mapping \cite{somma2002simulating} or some other exact mapping \cite{bravyi2002fermionic, seeley2012bravyi} will require $2N$ qubits.
The Hilbert space will be the same size, $4^N$, as expected, although the labelling of the states may look quite different, depending on the mapping. We will refer to this as using a qubit representation.

One is usually interested in simulating many-electron systems that satisfy one or more physical symmetries, and only a limited region of Hilbert space remains accessible. For example, a system with $N$ spatial orbitals with conserved particle number $n_\alpha + n_\beta = n_\text{elec}$ and projected spin $n_\alpha - n_\beta = 2S_z$, the relevant region of Hilbert space is spanned by $\mathcal{N}={N \choose n_\alpha} {N\choose n_\beta}$ basis states. The FQE is designed to take advantage of such symmetries to reduce the memory and CPU time associated with simulations of fermionic systems~\cite{rubin2021fermionic}.

Even accounting for symmetry, exact simulations of larger systems beyond roughly 18 spatial orbitals at half-filling are still infeasible. Describing a system with 18 spatial orbitals on a quantum computer requires 36 qubits, but larger system sizes are necessary to address practical questions associated with many algorithms for quantum simulation. To overcome the memory and CPU-time limitations associated with an exact description of the statevector, we utilize the MPS Ansatz for an approximate but systematically improvable representation of the fermionic wavefunction~\cite{white1999ab, chan2011density}.
The utility of the MPS Ansatz as an efficient representation %of the fermionic wavefunction 
has been rigorously proven for %describing 
the ground state of noncritical one-dimensional quantum systems with local interactions \cite{hastings2007area}. For more general quantum systems, the efficiency of the MPS Ansatz is not guaranteed, but high quality approximate MPS representations of the exact state are often found in practice that are more concise than the exact representation. Heuristic schemes that exploit the locality of interactions, in conjunction with orbital localization and reordering schemes, can be used to improve the compactness of the fermionic MPS Ansatz~\cite{wouters2014density}.

A general wavefunction may be written in MPS form as
\begin{equation}
    \ket{\Psi} = \sum_{\mathbf{n}} \mathbf{A}_1^{n_1}\mathbf{A}_2^{n_2}\ldots \mathbf{A}_N^{n_N}
    \ket{n_1, n_2, \ldots, n_N},
\end{equation}
where $n_i$ indexes the physical occupations of the $i$th orbital (or lattice site). Contracting the matrix dimensions (``virtual bonds'') of all the matrices for a given vector of physical occupations, $\mathbf{n} = (n_1, n_2, \ldots, n_N)$ will produce the amplitude associated with that basis state. This representation is clearly exact if the ``bond dimension" is allowed to be arbitrarily large. In fact, the MPS tensors can be obtained constructively by repeated singular value decomposition of the exact coefficient tensor. The dimension of the virtual bonds required for an exact representation grows exponentially toward the center of the tensor product so that the upper bound of the virtual bond summation for the $i$th tensor is obtained as $M_i = \min(4^i, 4^{L-i})$ for {an $L$-site system of} spin-1/2 fermions. As a result, in order to benefit from the MPS Ansatz, one must place a truncated upper bound on the bond dimension, $M$, so that $M_i = \min(4^i, 4^{L-i}, M)$. The effectiveness of the truncated MPS can be understood from information theoretic arguments~\cite{hastings2007area}, and it is known that the optimal approximate (truncated) wavefunction is obtained by retaining the states in a Schmidt decomposition with the largest Schmidt numbers~\cite{white1992density}. Note that construction of an MPS by consecutive SVDs of the FCI tensor does not yield the optimal truncated MPS wavefunction. Applying a local variational optimization to the truncated MPS Ansatz results in the density matrix renormalization group (DMRG) algorithm. Physical symmetries can be efficiently treated within the MPS representation by blocking the MPS tensors by {\it local} particle number and projected spin labels and then only storing those blocks which correctly multiply to the conserved {\it global} symmetry \cite{kurashige2009high}.

For efficient emulation of quantum algorithms, we must be able to apply unitary operators to emulate quantum gates and non-unitary operators to emulate projective measurement. In the MPS language, these operators can be expressed as matrix product operators (MPOs). Very generally, an operator can be written as a sum of products of operators acting on single sites,
\begin{equation}
    O = \sum_{\mathbf{z}} O^{z_1,z_2, \ldots, z_N} {o}_{z_1}{o}_{z_2}\ldots {o}_{z_N},
\end{equation}
where the index $z_i$ runs over a complete set of local operators. As with the MPS, one can represent this operator as a product of matrices (MPO):
\begin{equation}
    O = \sum_{\mathbf{z}} \mathbf{W}_1^{z_1}\mathbf{W}_2^{z_2}\ldots\mathbf{W}_N^{z_N}{o}_{z_1}{o}_{z_2}\ldots {o}_{z_N}.
\end{equation}
The bond dimension of the MPO is determined by the number of non-zero terms in the expansion and by the locality of the operators in each term. The efficient construction of MPOs of dense operators is a non-trivial problem, and the MPS-FQE can, via its interface with pyblock3, use either an SVD-based compression approach \cite{chan2016matrix} or the ``bipartite" approach \cite{ren2020general}. As with the MPS, the MPOs can be blocked by their symmetry labels. This is vitally important for efficient computation of expectation values and renormalized basis states used in the DMRG and time-dependent DMRG (td-DMRG) algorithms \cite{chan2016matrix}. Given a bra and ket MPS, the matrix element of the MPO can be exactly computed with an efficient contraction algorithm which is at the heart of the utility of DMRG. Unitary evolution of the MPS wavefunction is a more difficult problem and it has been an active research topic over the last 20 years. For the purposes of quantum emulation, we want to be able to apply the unitary operators of arbitrary, but generally sparse, generators. One option is to use the 4th order Runge-Kutta integrator (RK4),
\begin{align}
    \ket{\Psi(t + \Delta t)} &\equiv e^{-iH\Delta t}\ket{\Psi(t)} \nonumber\\
    &\approx \ket{\Psi(t)} + \frac{1}{6}\left[\ket{k_1} + 2\ket{k_2} + 2\ket{k_3} + \ket{k_4}\right],
\end{align}
where all of the $\ket{k_x}$ states can be obtained by only applying the generator, $H$. Here, and throughout the remainder of this manuscript, we have taken $\hbar = 1$. RK4 is conceptually simple, but it cannot be efficiently applied to the MPS wavefunction because exact application of the generator to a state causes the bond dimension to increase. Some kind of global or local compression must be performed after, or during, each application of the generator, but this will generally not be optimal for unitary evolution. This issue can be partially alleviated with algorithms that are designed for unitary evolution in the MPS representation \cite{cazalilla2002time,schmitteckert2004nonequilibrium,daley2004time,white2004real,feiguin2005time,zhao2008dynamics,ma2008dynamical,haegeman2011time,alvarez2011time,kinder2014analytic,haegeman2016unifying,ronca2017time,ma2018time,paeckel2019time}. In this work, we will primarily use the algorithm of Ronca {\it et al}~\cite{ronca2017time} which is a variant of the td-DMRG method of Feiguin and White~\cite{feiguin2005time}.

Importantly, all of the MPS algorithms discussed in this work can be symmetry adapted for particle number and projected spin. It is also possible to take advantage of the total spin squared (full SU(2)) symmetry within the DMRG/MPS framework \cite{mcculloch2002non,sharma2012spin,wouters2014density}, but this symmetry is not currently supported by the FQE interface. Using the local particle number blocks, it is also simple to treat the signs that arise from interchanging fermions arising from the application of operators. Formally, our MPS Ansatz can be considered as a fermionic MPS~\cite{cirac2021matrix}. 

\begin{figure}
    \centering
    \includegraphics[width=\linewidth]{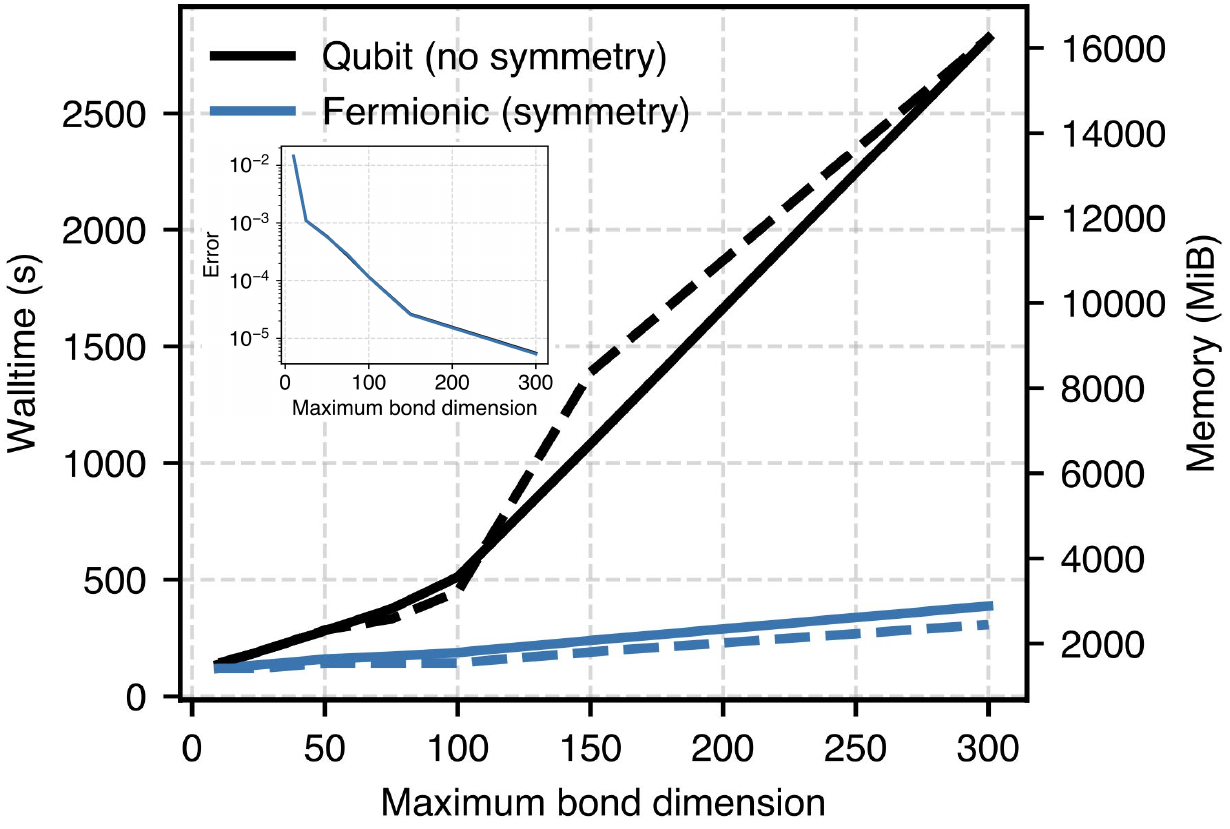}
    \caption{Comparison of CPU time (solid lines) and memory (dashed lines) for simulating the application of a layer of one-electron gates to a system with 14 sites and 14 electrons over a range of bond dimensions. The fermionic representation results utilize particle number and projected spin symmetry, while the qubit representation results do not. As shown in the inset, the error compared with exact results is nearly {graphically indistinguishable between} the two representations. {Both simulations were performed using the ITensors.jl library~\cite{ITensor,ITensor-r0.3}. A layer of 91 one-electron gates with randomly-sampled coefficients was applied to an initial Hartree--Fock state. The simulations were performed with 8 threads on Intel Cascade Lake cores using block sparse threading and the reported timing does not include garbage collection. The reported memory corresponds to the maximum amount of RAM consumed at any point along to simulation. The garbage collector was called after each gate was applied using a single RK4 step. The error was computed as the norm of the difference between the computed and reference one particle density matrices, normalized by the number of sites.}}
    \label{fig:qubit_vs_fermion}
\end{figure}

Using a symmetry-adapted fermionic MPS wavefunction can be significantly more efficient than using an MPS wavefunction without symmetry.
In Figure~\ref{fig:qubit_vs_fermion} we show the advantage of symmetry explicitly by applying a complete layer of one-electron, orbital-rotation gates to an initial mean-field (product state) MPS wavefunction. The total wall time and peak memory usage are shown as a function of bond dimension in the symmetry-adapted fermionic representation ($N$ sites, 4 occupation states per site, using particle number and projected spin symmetry) and in the non-symmetry adapted qubit represention ($2N$ sites, 2 states per site without symmetry restriction, using the Jordan-Wigner mapping). The error due to the bond dimension truncation is shown in the inset of Figure~\ref{fig:qubit_vs_fermion} and it is largely independent of the representation and the use of symmetry. Therefore, for simulations that satisfy one or more symmetries, it is clearly superior to use a representation that allows for an efficient, straightforward treatment of symmetry. The software presented in this work implements the FQE interface with an MPS backend that fully leverages particle number and projected spin symmetry in a manner consistent with the behavior of the FQE.

\section{Software implementation}\label{sec:Implementation}
The key contribution in the described work is the implementation of an MPS backend to the FQE that greatly expands the size of quantum circuits that can be emulated. The described MPS backend utilizes the pyblock3 library~\cite{pyblock3} for performing basic tensor operations and, when possible, the block2 library~\cite{zhai2023block2} for optimal performance in a parallel computing environment.
The MPS backend offers the ability to substitute the exact statevector with a memory-efficient approximation that is systematically improvable. Crucially, one can often still obtain ``exact'' results for observable quantities \textit{via} extrapolation of the maximum bond dimension. 

The MPS-FQE emulator includes class methods to simplify integration of the MPS backend into existing FQE-based emulation workflows. For example, one can generate an \code{MPSWavefunction} object, the central object for manipulation in MPS-FQE, from the FQE \code{Wavefunction} object with minimal effort (or vice versa), and the consistency between the two APIs enables the \code{MPSWavefunction} to function as nearly a drop-in replacement for its FQE counterpart. When generating an instance of the \code{MPSWavefunction}, one has the option to specify MPS-specific parameters, or else inherit the corresponding default values, that influence the behavior of subsequent manipulations of the wavefunction. For example, the maximum bond dimension and the cutoff value used for compression and truncation of MPS tensors can be specified upon instantiating the object, or it may be specified explicitly upon calling methods of the \code{MPSWavefunction} object.
\begin{figure}[h!]
    \centering
    \includegraphics[width=\linewidth]{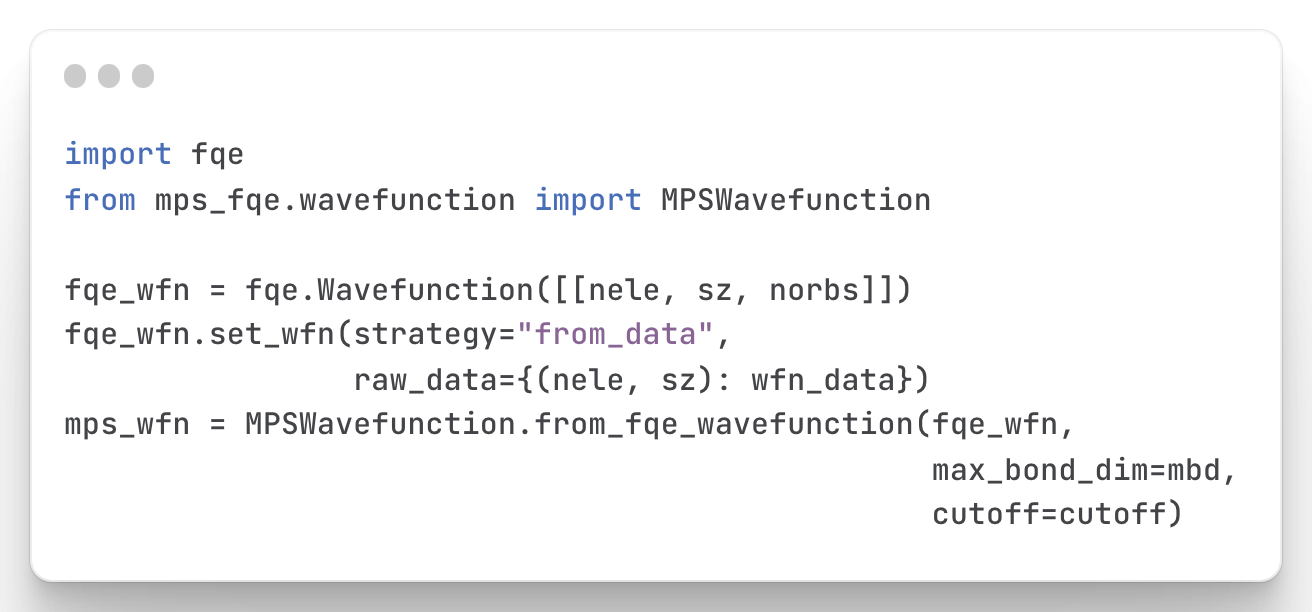}
\end{figure}

Moreover, manipulations of the \code{MPSWavefunction} involving application of operators invoke a flexible interface that allows users to provide FQE operators as arguments in addition to MPOs. When dealing with large systems and high bond dimension, it may be prohibitively costly to apply operators in an exact manner through tensor contractions, even when performing subsequent compression. To alleviate this issue, MPS-FQE offers the option to utilize an approximate sweep algorithm for operator application with a reduced memory demand \cite{schollwock2011density, zhai2023block2}.
\begin{figure}[h!]
    \centering
    \includegraphics[width=\linewidth]{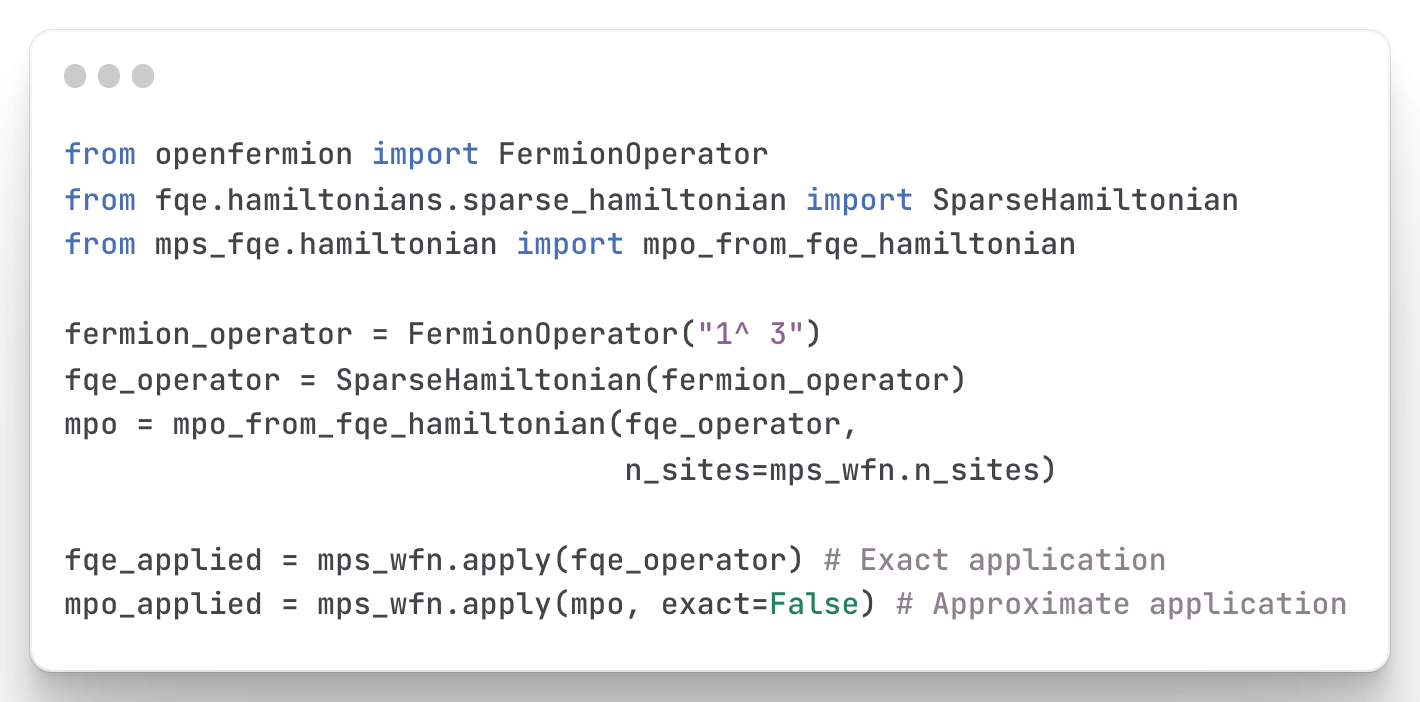}
\end{figure}

Finally, a variety of time evolution algorithms are available including the td-DMRG and RK4 with both exact and approximate application of the generator. The user has the option to include keywords that specify the details of the chosen time evolution algorithm, or else inherit default values. 
\begin{figure}[h!]
    \centering
    \includegraphics[width=\linewidth]{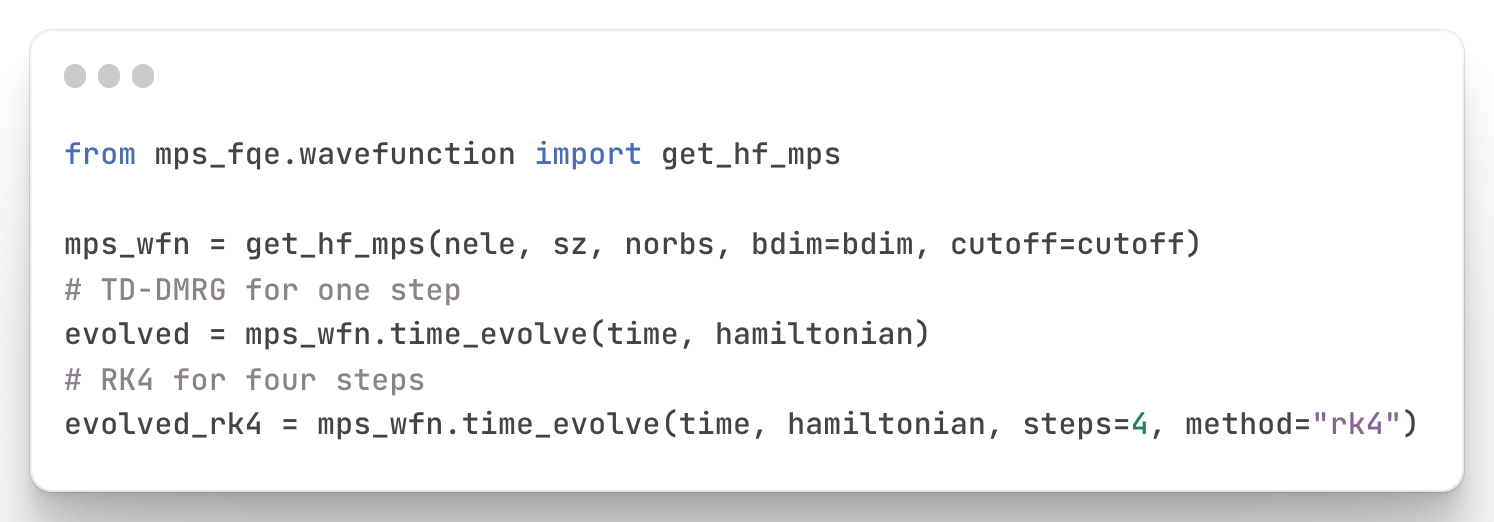}
\end{figure}

\section{Applications}\label{sec:Applications}
Here we describe four potential applications of the MPS-FQE that are of particular relevance to current research directions in quantum algorithms for simulating fermionic systems. In each case we show that MPS-FQE can provide useful, near-exact results for systems where exact state vector emulation is impossible.

\subsection{Preparing near-exact ground states}
\begin{figure}
    \centering
    \includegraphics{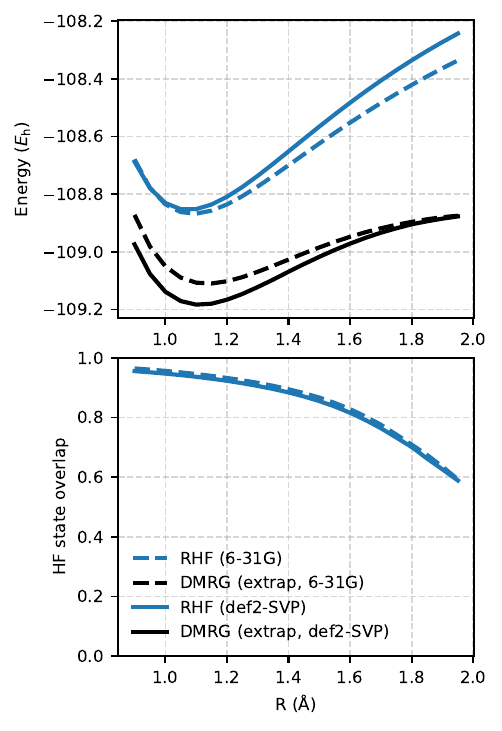}
    \caption{Comparison of Hartree--Fock and exact potential energy curves for $\mathrm{N}_2$ (top) and the overlap between the exact state and the Hartree--Fock state (bottom). Results are shown in the 6-31G basis set (14 electrons in 18 spatial basis functions) and the def2-SVP basis set (14 electrons in 28 spatial basis functions). All ``exact'' quantities are obtained by extrapolating DMRG results {from $M=400$ to zero truncation error.} Analogous exact calculations would require approximately 15 GB and 22 TB respectively to store a single state vector even accounting for particle number and spin symmetry.}
    \label{fig:N2_dmrg}
\end{figure}

Many quantum algorithms for the ground state of fermionic systems use some variant of quantum phase estimation (QPE) which requires a significant overlap between the initial state, $\ket{\Psi_0}$, and the exact ground state, $\ket{\Psi}$. MPS-FQE can be used both as a means to approximate the exact ground state, $\ket{\Psi}$, and as a way to emulate state preparation algorithms for $\ket{\Psi_0}$. In the former case, DMRG is the most efficient way to form near-exact states in the MPS representation, and a seamless interface with the pyblock3 and block2 codes makes it easy to convert the DMRG solution to an MPS-FQE wavefunction. For example, in Figure~\ref{fig:N2_dmrg} we show the potential energy and the overlap, {$\left|\braket{\Psi_0 | \Psi}\right|$}, for a restricted Hartree--Fock (RHF) initial state of N$_2$. The ``exact'' quantities are computed from extrapolating DMRG results to the limit of infinite bond dimension. In this particular example, the overlap between the RHF state and the ``exact" ground state is large enough such that the RHF state could be used for QPE over most of the potential energy surface. Note that it is useful to be able to compute these quantities for larger basis sets, as we do here, because minimal basis sets that are often used while discussing quantum algorithms are too small to provide quantitative information with chemically-meaningful accuracy. The calculations with the 6-31G basis set are still possible using an exact statevector solver, but the def2-SVP basis is far too large for an exact solution. 

\begin{figure}
    \centering
    \includegraphics{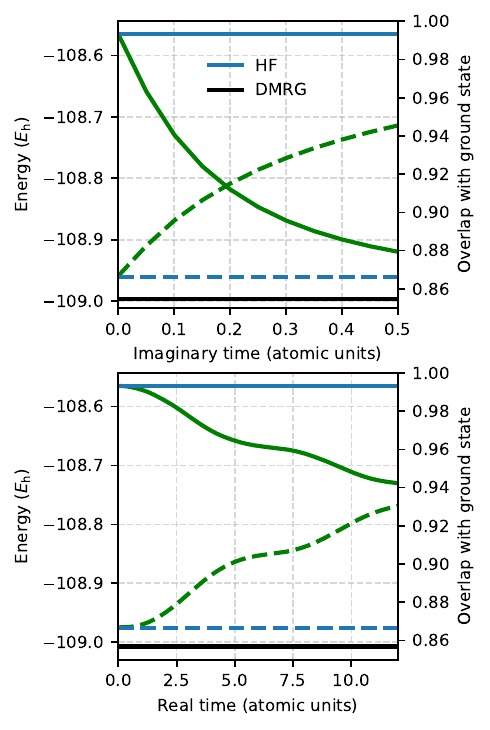}
    \caption{A comparison of imaginary time evolution (top) and adiabatic state preparation (bottom) for N$_2$ ($R = 1.5 \text{\AA}$) in the def2-SVP basis set {(green lines)}. Results are shown for a max bond dimension of $M = 200$ where both simulations start from the HF ground state, and energy (solid lines) and overlap with the exact ground state (dashed lines) are shown. The total evolution time of the adiabatic state preparation with a linear schedule is chosen to be 20 atomic units. Even accounting for symmetries, exact emulation of these algorithms would require more than 22 TB of memory for this system.}
    \label{fig:N2_state_prep}
\end{figure}

MPS-FQE can also be used to emulate state preparation algorithms on quantum computers. For example, one might want to use a small amount of quantum imaginary time evolution or adiabatic state preparation to prepare a state that has stronger overlap with the exact ground state. In this case, the qualitative information provided by MPS-FQE with a modest bond dimension can be very useful. In Figure~\ref{fig:N2_state_prep} we compare the amount of imaginary and real time respectively necessary to prepare a state where the overlap with the exact ground state is greater than 0.9. {Note that the ``exact" ground state is estimated by extrapolating DMRG calculations to zero truncation error from $M=400$.} Again we show results for N$_2$ ($R = 1.5 \text{\AA}$) in the def2-SVP basis set where the 28 spatial orbitals (56 spin orbitals) makes exact emulation  impossible. Although imaginary time evolution is much more efficient in terms of the {number of time steps, each step of imaginary time is much more complicated to implement} on a quantum computer~{\cite{motta2020determining, hejazi2023adiabatic}} and the most efficient algorithm will depend on the implementation on real hardware. Either way, MPS-FQE can provide useful estimates {for} such applications.

\subsection{Testing VQE Ans\"atze}
Designing efficient, low-depth circuits and optimizing the parameters of those circuits is the primary challenge of hybrid algorithms. During the NISQ era, hybrid algorithms are some of the most practical candidates for evaluating
real-world applications of quantum computation. The Variational Quantum Eigensolver (VQE) is a popular hybrid algorithm designed to report a variational upper bound on the ground state energy of a quantum system  
\cite{peruzzo2014variational, mcclean2016theory,tilly2022variational}.

Since its introduction, several VQE Ans\"atze have been proposed \cite{romero2018strategies, lee2018generalized, grimsley2019adaptive, tilly2022variational}, and efficient emulation of such quantum algorithms can be useful for exploring novel fermionic gate primitives or for evaluating the difficulty of classically simulating a quantum circuit. In particular, investigating more realistic systems with larger basis sets is important when exploring the practical issues such as convergence under different classical optimization schemes. As an example, we consider the popular ADAPT-VQE ansatz~\cite{grimsley2019adaptive} and show how MPS-FQE can be used to emulate the algorithm applied to larger chemical systems.

The ansatz obtained {via} the ADAPT-VQE algorithm is, in principle, exact for a given basis~{\cite{grimsley2019adaptive}} and can be expressed {to $N$th order} as
\begin{equation}\label{eq:adapt}
    \ket{\Psi}_{\text{ADAPT}} = e^{-\theta_N{O}_N} e^{-\theta_{N-1}{O}_{N-1}} \dots e^{-\theta_1{O}_1} \ket{\Psi_\text{HF}}.
\end{equation}
Here, $e^{-\theta_i{O}_i}$ is a parametrized unitary and $\ket{\Psi_\text{HF}}$ is the Hartree--Fock state. The operators, $\{{O}_i\}$, are chosen based on the magnitude of the energy gradient from an ``operator pool'' that contains the set of spin-conserving, generalized single and double excitation operators~\cite{grimsley2019adaptive}. The ansatz expands in an iterative fashion where, during each iteration, the parameters $\{\theta_i\}$ are re-optimized and a new (possibly repetitive) operator is selected from the pool based on the updated energy gradient. 

\begin{figure}
    \centering
    \includegraphics[width=\linewidth]{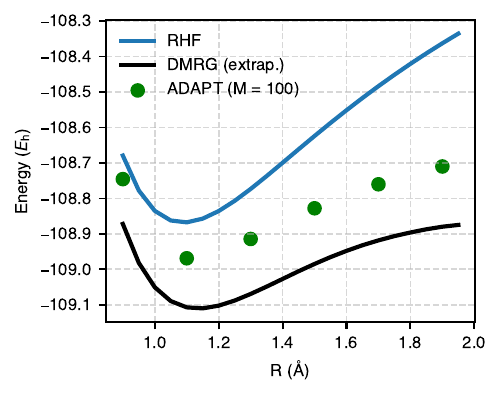}
    \caption{MPS-FQE emulation of ADAPT-VQE for N$_2$ in the 6-31G basis set. The ``exact" ground state energy is obtained from DMRG extrapolated to zero truncation error.}
    \label{fig:adapt}
\end{figure}

In Figure~\ref{fig:adapt}, we present ADAPT-VQE results obtained with MPS-FQE for N$_2$ using the 6-31G basis set with a max bond dimension of $M=100$. For the calculations described here, we have evaluated the operator pool gradients using exact analytical commutation relations (although all Hamiltonian applications are performed using an approximate sweep algorithm) despite the fact that any unitary corresponding to a selected operator is ultimately applied approximately \textit{via} RK4 followed by compression under the imposed maximum bond dimension and a cutoff value of $10^{-20}$. The ADAPT ansatz is subsequently optimized at each iteration using analytical commutation relations. We also note that, with respect to the exact potential energy curve, one obtains significantly better agreement simply by performing DMRG simulations, representing a direct variational optimization of the MPS ansatz. The ADAPT-VQE ansatz, on the other hand, may be much more sensitive to orbital ordering given the nature of the applied excitation operators. In fact, we have performed the same set of calculations with $M=200$ and observed only a very small (almost graphically indistinguishable) improvement to the energy. While we have not utilized any orbital localization or reordering schemes in this application, we note that the MPS-FQE library is well-suited for evaluating their impact in this context, and it is reasonable to expect that the use of such schemes would improve approximate MPS-based ADAPT-VQE results.

While a conventional ADAPT-VQE implementation considers convergence with respect to the $L^2$ norm of the operator pool gradient vector \cite{grimsley2019adaptive}, we have found in this application that it can be challenging to converge the gradient norm below about $2\times10^{-1}$ in many cases (compared to $10^{-1}$ in the ADAPT($\varepsilon_1$) implementation) using the described scheme for evaluating gradients with respect to parameters for operators in the pool. The reason the operator pool gradient norm remains relatively large in some cases is likely due to the mismatch between the manner in which the gradients are evaluated and the selected unitaries are applied. The operator pool gradient vector is computed assuming that the application of each corresponding unitary will be done exactly (\textit{i.e.} they are obtained via the analytical commutation relations). The quality of this assumption likely deteriorates as the ADAPT ansatz grows and the impact of a truncated bond dimension becomes more severe for larger iterations. Thus, it remains possible that an operator selected from the pool, which is expected to have the greatest influence on the energy under the assumption of an exact application of the corresponding unitary, may have a considerably smaller impact under truncated bond dimension.

An alternative, and perhaps more internally consistent, implementation would evaluate all gradients \textit{via} finite differences, such that they reflect the approximate nature of propagation in a truncated Hilbert space. However, evaluating the gradients of every term in the expansive operator pool \textit{via} finite differences would be prohibitively expensive for the basis set considered here. Analytic evaluation of the gradient of the approximate ansatz is also possible in principle, but the implementation is difficult, requiring the solution of response equations at each step. The data shown in Figure~\ref{fig:adapt} therefore corresponds to the endpoint of an ADAPT trajectory that is terminated when the cumulative change in energy over 3 successive iterations is less than $10^{-5} \ E_\text{h}$ or, in some cases, when the gradient norm converges below $10^{-1}$.

This emulation of ADAPT-VQE contains error from the approximate nature of the ADAPT-VQE method, error due to the truncated bond dimension, and error due to the adopted convergence scheme. The former includes the error due to the finite number of iterations used to optimize the ansatz as well as the possibility of converging to a local minimum. Unfortunately, these sources of error are coupled, which makes extrapolation to the ``exact" ADAPT-VQE energy difficult. For example, the ADAPT-VQE solution at $M = 100$ may be a different solution than the one found at $M = 200$. Despite these difficulties, MPS emulation in the fermionic representation allows one to study the qualitative features of the ansatz for much larger systems. In this case, the basis of 36 spin orbitals is much larger than systems studied with ADAPT-VQE in the past.

\subsection{Numerical evaluation of Trotter error}
When doing QPE or simulating any kind of fermionic dynamics on a quantum computer, one way to apply the many-body unitary operator is to use an approximate product formula where the unitary generated by a sum of terms is approximated as a product of unitary operators. The simplest such formula is the Lie--Trotter product formula,
\begin{equation}
    e^{-i(A + B)\Delta t} = e^{-iA\Delta t}e^{-iB\Delta t} + O(\Delta t^2),
\end{equation}
which has an error that is quadratic in the propagation time, $\Delta t$, with a coefficient that is proportional to the norm of the commutator between $A$ an $B$. 

To decide on the most efficient strategy for real applications, an asymptotic understanding of the error is not sufficient and the coefficients must be also considered. For example, for the simple Lie--Trotter formula,
\begin{equation}
\begin{split}
    ||e^{-i(A + B)\Delta t}& - e^{-iA\Delta t}e^{-iB\Delta t}||_2  = \\ 
    &  \frac{1}{2}||[A,B]||_2 \Delta t^2 + O(\Delta t^3).
\end{split}
\end{equation}
The 2-norm, or spectral norm, of an operator can be estimated with power iteration, and this can be accomplished with MPS-FQE.

\begin{figure}
    \centering
    \includegraphics{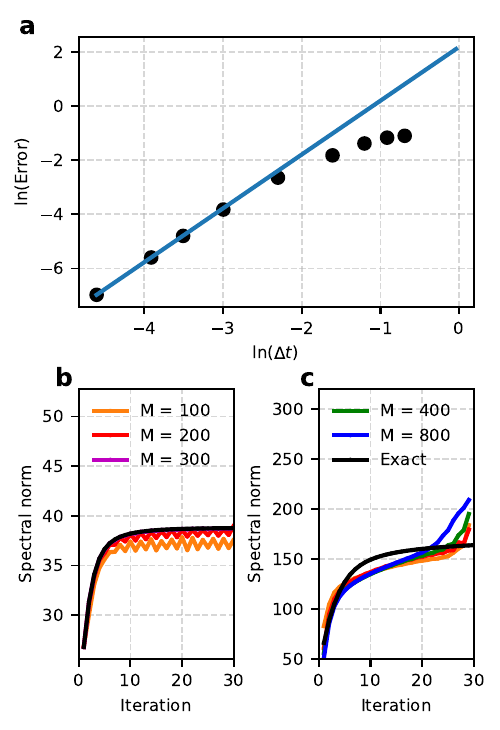}
    \caption{The estimation of Trotter error with FQE and MPS-FQE. (a) log-log plot of $||e^{-i(T + V)\Delta t} - e^{-iT\Delta t}e^{-iV\Delta t}||_2$ vs the time step in atomic units (black dots) and the expected asymptotic behavior (blue) computed from $\frac{1}{2}||[T,V]||_2$ for a system of 4 electrons in a grid basis of 27 orbitals {computed with FQE}. The spectral norm is computed with power iteration. The spectral norm of the commutator (from power iteration) is shown with exact {(FQE)} and truncated bond dimension {(MPS-FQE)} for (b) 4 electrons in 27 orbitals and (c) 4 electrons in 64 orbitals.}
    \label{fig:trotter}
\end{figure}

As an example, we consider the electron gas Hamiltonian recently used to obtain quantum resource estimates for stopping power calculations \cite{rubin2023quantum}, which has the form,
\begin{align}
    H &= T + V \nonumber \\ 
    &= \sum_{jk,\sigma} \tau_{jk}a^{\dagger}_{j\sigma}a_{k\sigma} + \sum_{lm,\sigma\tau}\nu_{lm}a^{\dagger}_{l\sigma}a_{l\sigma}a^{\dagger}_{m\tau}a_{m\tau}.
\end{align}
Here $\sigma$ and $\tau$ index the spin, and $T$ and $V$ refer to the kinetic and potential energy respectively. It is known from reference~\cite{PRXQuantum.4.020323} that, for an operator splitting into kinetic energy and potential energy terms, the spectral norm of the difference between the exact and trotterized propagators has scaling proportional to polynomials of the particle-weighted norms of each operator,
\begin{equation}\label{eq:Low_bound}
\begin{split}
||&e^{-iH\Delta t} - S_k(\Delta t)||_2 = \\
& \mathcal{O}\left((\norm{\tau}_{1}+\norm{\nu}_{1,[\eta]})^{k-1}\norm{\tau}_1\norm{\nu}_{1,[\eta]}\eta\, \Delta t^{k+1}\right)
\end{split}
\end{equation}
where
\begin{align}
\|\tau\|_{1} &= \max_{j} \sum_{k}|\tau_{j,k}| \\
\|\nu\|_{1,\left[\eta\right]} &= \max_{j} \max_{k_{1} < ... < k_{\eta}} \left(|v_{j,k_{1}}| + ... + |v_{j, k_{\eta}}| \right),
\end{align}
$k$ is the Trotter order, and $\eta$ is the number of electrons. Thus in order to determine the constant factors we must determine the true spectral norm of the difference between the exact operator and the product formula and divide by the norm based complexity inside the big-O of Eq.~\eqref{eq:Low_bound}. Since the constant factors themselves are not relevant to this work, we will describe only the computation of the spectral norm that appears on the LHS of Equation~\ref{eq:Low_bound} (see reference~\cite{PRXQuantum.4.020323} for a more complete discussion of how to compute and interpret the constant factors). In Figure~\ref{fig:trotter}, we show how one can estimate coefficients in the error incurred by using the Lie--Trotter formula. We note that this problem is not ideally suited to approximation with matrix product states because computation of the 2-norm requires us to compute the state whose norm is maximized by the action of an operator. Such states are not generally low-entanglement, and this makes it difficult to accurately perform the power iteration to compute the 2-norm under truncated bond dimension. Furthermore, the approximate and state-dependent application of the operators means that it is possible for power iteration to collapse to an eigenvalue that is much larger than those of the exact operator. This behavior can be seen in Figure~\ref{fig:trotter}, (b) and (c) where it is particularly obvious for the calculations done with small bond dimension. Despite these limitations, qualitative information can still be obtained with a low-bond-dimension MPS, and quantitative information can be obtained with a lower memory footprint than that of the exact calculation.

\subsection{Near-exact simulation of fermion dynamics}
Many interesting and important properties of many-body fermionic systems are nonequilibrium in nature, and are determined by the transient behavior of the system following interactions with external stimuli. Understanding the dynamics of nonequilibrium quantum systems is crucial for the development of future technologies and it is expected that simulating the coherent nonequilibrium dynamics of interacting quantum systems will be a relevant application of future FTQC devices~\cite{polkovnikov2011colloquium}. For systems beyond the size of those that are exactly treatable, the MPS-FQE simulator offers the ability to explore quantum dynamical behavior in an approximate, yet systematically improvable, manner. For demonstration purposes, we consider the dynamics of a 20-site Fermi-Hubbard chain at half-filling under periodic boundary conditions.

The Fermi-Hubbard model is known to be an effective proxy for strongly correlated fermionic systems, and it contains much of the essential physics necessary to describe many of their exotic properties despite its relative simplicity~\cite{anderson1987resonating, cocchi2016equation}. The Hamiltonian for the one-dimensional Fermi-Hubbard chain is given by
\begin{equation}
{H} = -t_h \sum_{i, \sigma} \left( {a}^\dagger_{i, \sigma}{a}_{i+1, \sigma} + \text{h.c.}\right) + U \sum_{i} {n}_{i,\uparrow} {n}_{i,\downarrow}.
\end{equation}
Here, $t_h$ is the nearest-neighbor tunneling amplitude, $U>0$ is the on-site electron-electron repulsion, and ${n}_{i,\sigma} = {a}^\dagger_{i, \sigma}{a}_{i, \sigma}$ is the fermionic number operator.

\begin{figure}
    \centering
    \includegraphics[width=\linewidth]{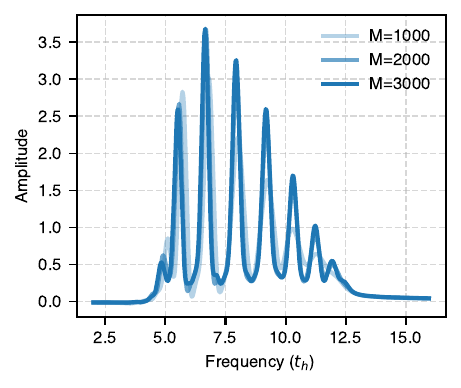}
    \caption{Ground state optical spectrum for a 20-site Hubbard chain in the strongly interacting regime ($U=8 t_h$) from $M=1000$ to $M=3000$.}
    \label{fig:Optical}
\end{figure}

In Fig.~\ref{fig:Optical}, we report the ground state optical spectrum obtained as the temporal Fourier transform of the two-point current operator autocorrelation function defined for a chain of length $L$ as
\begin{equation}
A(\omega) \propto \frac{1}{L} \int_{0}^{\infty} d\tau \ e^{i(\omega + i \gamma)\tau} \text{Tr}[{J} e^{-i({H}-E_0)\tau}{J} {\rho}_0],
\end{equation}
where ${J} = -i t_h \sum_{i, \sigma} \left({a}^\dagger_{i,\sigma}{a}_{i+1,\sigma} - \text{h.c.}\right)$ is the current operator, ${\rho}_0 = \ket{\psi_0}\bra{\psi_0}$ is the initial density operator, and $\gamma = 0.1 \ t_h^{-1}$ is a Lorentzian broadening factor~\cite{maldague1977optical}. From the above expression, it is clear that the peak structure of the ground state optical spectrum is determined by oscillatory behavior of coherence elements between the ground state, $\ket{\psi_0}$ and the excited states that are prepared {via} the action of the current operator. The simulations reported here were performed up to $t_\text{max} = 20 \ t_h^{-1}$, and we have included a cosine window function in the Fourier integral to ensure that the (damped) correlation function is identically 0 at $\tau = t_\text{max}$. For this one-dimensional system, the spectrum is fully converged at a maximum bond dimension of $M=3000$, as it is graphically indistinguishable from one obtained with $M=3500$, but a qualitatively correct spectrum can be obtained with much lower bond dimensions as shown in Figure~\ref{fig:Optical}. This application, which involves calculating the initial state, $\ket{\psi_0}$, with DMRG and then subsequently calculating real-time dynamics of the system upon a quench (here, obtained by applying the current operator), highlights the broad range of problems that the MPS-FQE emulator is capable of treating.

\section{Conclusions}
In summary, the matrix product state backend for FQE (MPS-FQE) can serve as a drop-in replacement for FQE, and the controllable approximation of the MPS representation enables the treatment of problems that are much larger than those accessible with an exact statevector emulator. Like FQE, MPS-FQE is designed to work in the standard fermionic representation and fully supports particle number and projected spin symmetries. The code is available on github under the GPL-3.0 license \cite{mps_fqe_2023}. In Section~\ref{sec:Applications} we have described several examples of the types of quantum simulation problems where FQE, and by extension MPS-FQE, can be useful. For these problems, MPS-FQE can lower the cost of quantum emulation to allow calculations on larger, more realistic systems.

Possible future work includes the implementation of automatic differentiation and extensions appropriate for different parallel environments including graphical processing units (GPUs). One weakness of the current version of MPS-FQE is that, unlike FQE, it does not yet have specialized routines for unitary evolution with structured generators. For example, when the generator is very sparse, the general time evolution algorithm may perform more work than is strictly necessary. Specialized algorithms to take advantage of the structure of common generators in the MPS represention would be valuable future addition to the library. 

\section{Acknowledgments}
Part of the work performed at Quantum Simulation Technologies, Inc. was supported by the US National Science Foundation under Grant No.  2126857. Development of block2 and pyblock3 was primarily supported by the US National Science Foundation,
under Award No. CHE-2102505.

\bibliography{references}
\end{document}